\pdfoutput=1

\documentclass[11pt]{article}

\usepackage[]{acl}

\usepackage{times}
\usepackage{latexsym}
\usepackage{microtype}
\usepackage{graphicx}
\usepackage{booktabs}  
\usepackage{multirow}
\usepackage{subfigure}
\usepackage{verbatim}
\usepackage{CJKutf8} 
\usepackage{amssymb}
\usepackage{amsmath}

\usepackage{algorithm}
\usepackage{algorithmic}
\usepackage{makecell}
\usepackage[T1]{fontenc}

\usepackage[utf8]{inputenc}

\usepackage{microtype}

\title{Two Birds with One Stone: Unified Model Learning for Both Recall and Ranking in News Recommendation}

\author{Chuhan Wu$^\dagger$~~~~Fangzhao Wu$^{\ddagger}$\thanks{~~Corresponding author.}~~~~Tao Qi$^\dagger$~~~~Yongfeng Huang$^\dagger$\\
    $^\dagger$Department of Electronic Engineering, Tsinghua University, Beijing 100084, China  \\
     $^\ddagger$Microsoft Research Asia, Beijing 100080, China\\
  {\tt\{wuchuhan15, wufangzhao, taoqi.qt\}@gmail.com} \\ {\tt yfhuang@tsinghua.edu.cn}
  }

\begin{document}
\maketitle
\begin{abstract}

Recall and ranking are two critical steps in personalized news recommendation.
Most existing news recommender systems conduct personalized news recall and ranking separately with different models.
However, maintaining multiple models leads to high computational cost and poses great challenges to meeting the online latency requirement of news recommender systems.
In order to handle this problem, in this paper we propose UniRec, a unified method for recall and ranking in news recommendation.
In our method, we first infer user embedding for ranking from the historical news click behaviors of a user using a user encoder model.
Then we derive the user embedding for recall from the obtained user embedding for ranking by using it as the attention query to select a set of basis user embeddings which encode different general user interests and synthesize them into a user embedding for recall.
The extensive experiments on benchmark dataset demonstrate that our method can improve both efficiency and effectiveness for recall and ranking in news recommendation.
\end{abstract}
\section{Introduction}

\begin{figure}[!t]
  \centering
    \includegraphics[width=0.99\linewidth]{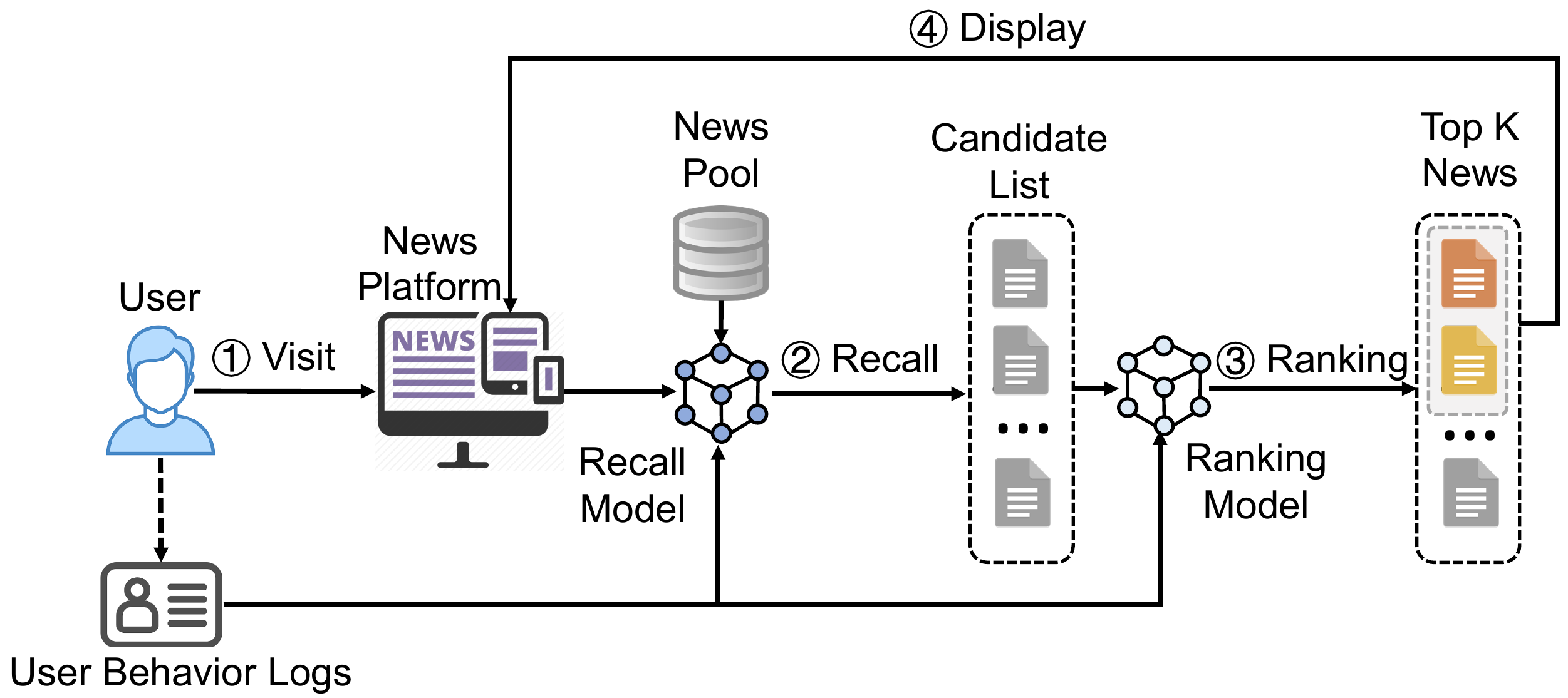}

  \caption{A typical pipeline of news recommendation.}
  \label{fig.pipeline}
\end{figure}

News recommendation techniques are widely used by many online news websites and Apps to provide personalized news services~\cite{wu2020mind}.
Recall and ranking are two critical steps in personalized news recommender systems~\cite{karimi2018news,wu2021personalized}.
As shown in Fig.~\ref{fig.pipeline}, when a user visits a news platform, the recommender system first recalls a set of candidate news from a large-scale news pool, and then ranks candidate news for personalized news display~\cite{wu2020mind}. 
Both news recall and ranking have been widely studied~\cite{elkahky2015multi,liu2019hi,liu2020octopus,wu2020user,wang2020fine,wu2021feedrec,qi2021personalized,qi2021pp,qi2021uni,qi2021hierec}.
In online news recommender systems, recall and ranking are usually conducted separately with different models, as shown in Fig.~\ref{fig.pipeline}.
However, maintaining separate models for news recall and ranking in large-scale news recommender systems usually leads to heavy computation and memory cost~\cite{tan2020learning}, and it may be difficult to meet the latency requirement of online news services.

Learning a unified model for personalized news recall and ranking would be greatly beneficial for alleviating the computation load of news recommender systems.
However, it is a non-trivial task because the goals of recall and ranking are not the same~\cite{covington2016deep,malkov2018efficient}.
Ranking usually aims to accurately rank candidates based on their relevance to  user interests~\cite{wu2019npa,ge2020graph,wu2021user,wang2020fine}, while recall mainly aims to form a candidate pool that can comprehensively cover user interests~\cite{liu2020octopus,qi2021hierec}.
Thus, the model needs to adapt to the different goals of recall and ranking without hurting their performance.

In this paper, we propose a news recommendation method named UniRec, which can learn a unified user model for personalized news recall and ranking. 
In our method, we first encode news into embeddings with a news encoder, and learn a user embedding for ranking from the embeddings of historical clicked news.
We further derive the user embedding for recall by using the user embedding for ranking as the attention query to select a set of basis user embeddings that encode different general user interest aspects and synthesize them into a user embedding for recall.
In the test phase, we only use the basis user embeddings with top attention weights to compose the  user embedding for recall to filter noisy user interests.
Extensive experiments on a real-world dataset demonstrate that our method can conduct personalized news recall and ranking with a unified model and meanwhile achieve promising recall and ranking performance.

\begin{figure}[!t]
  \centering
    \includegraphics[width=0.99\linewidth]{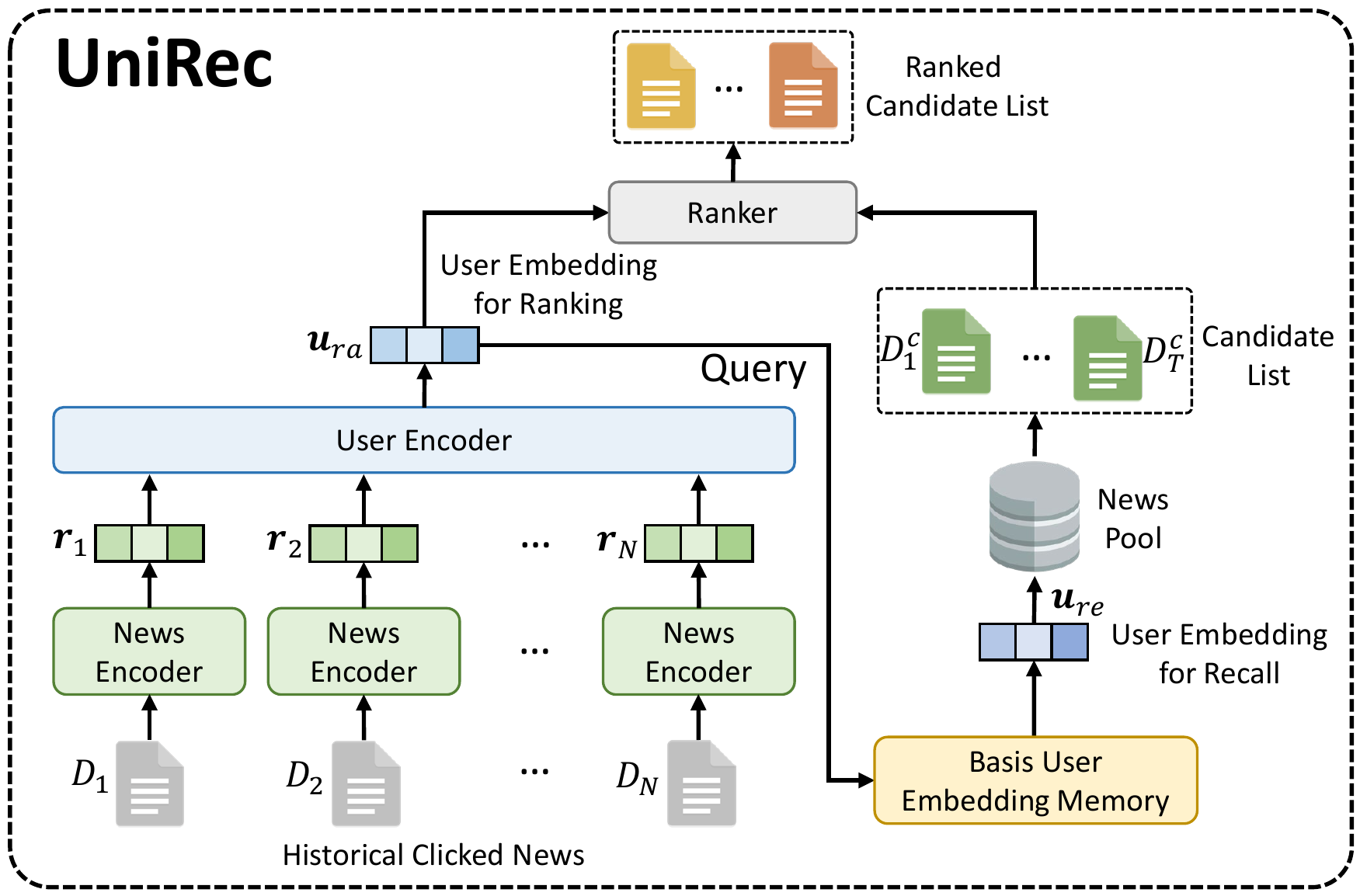}
  \caption{The framework of \textit{UniRec}.}
  \label{fig.model}
\end{figure}

\section{Methodology}\label{sec:Model}

The overall framework of \textit{UniRec} is shown in Fig.~\ref{fig.model}.
We first learn a user embedding for ranking from  the user's historical clicked news.
We then derive a user embedding for recall from the user embedding for ranking and a set of basis user embeddings that encode different general  interests.
Their details are introduced as follows.

\subsection{Ranking for News Recommendation}

The ranking part aims to rank candidate news in a small candidate list according to user interests.
Following~\cite{wu2020mind}, \textit{UniRec} uses a news encoder that learns news embeddings from news texts and a user encoder that learns user interest embedding for ranking from the embeddings of clicked news.
The candidate news embedding and user  embedding for ranking are used to compute a click score for personalized news ranking.
More specifically, we denote a user $u$ has $N$ historical clicked news $[D_1, D_2, ..., D_N]$.
These clicked news are encoded into a sequence of news embeddings, which is denoted as $[\mathbf{r}_1, \mathbf{r}_2, ..., \mathbf{r}_N]$.
The user encoder further takes this sequence as input, and outputs a user embedding  $\mathbf{u}_{ra}$ for ranking.
For a candidate news $D^c_i$, we use the news encoder to obtain its embedding $\mathbf{r}^c_i$.
We follow~\cite{okura2017embedding} to compute the probability score of the user $u$ clicking on the candidate news $D^c_i$ via inner product, i.e., $\hat{y}_{ra}^i=\mathbf{u}_{ra} \cdot \mathbf{r}^c_i$.
The click scores of the news in a candidate list are used for personalized ranking.
Following~\cite{wu2019nrms}, we use multi-head self-attention networks in both news and user encoders to capture the contexts of words and click behaviors, respectively.
In addition, following~\cite{devlin2019bert} we add position embeddings to capture the orders of words and behaviors.

\subsection{Recall for News Recommendation}

The recall part aims to select candidate news from a large news pool based on their relevance to user interests.
To efficiently exploit user interest information for personalized news recall, we take the  user  embedding for ranking as input instead of rebuilding user interest representations from original user click behaviors.
However, since the goals of ranking and recall are not the same~\cite{kang2019candidate}, the user embedding for ranking may not be suitable for news recall.
Thus, we propose a method to distill a user embedding for recall from the user embedding for ranking.
More specifically, we maintain a basis user embedding memory that encodes different general  user interest aspects.
We denote the $M$ basis user embeddings in the memory as $[\mathbf{v}_1, \mathbf{v}_2, ..., \mathbf{v}_M]$.
We use the user embedding for ranking as the attention query to select basis user embeddings.
We denote the attention weight of the $i$-th basis user embedding as $\alpha_i$, which is computed as:
\begin{equation}
    \alpha_i= \frac{\exp(\mathbf{u}_{ra}\cdot \mathbf{w}_i)}{\sum_{j=1}^M \exp(\mathbf{u}_{ra}\cdot \mathbf{w}_j)},
\end{equation}
where the parameters $\mathbf{w}_i$ are served as the attention keys.
Different from additive attention~\cite{yang2016hierarchical} where the attention keys and values are equivalent, in our approach the keys (i.e., $\mathbf{w}_i$) are different from the values (i.e., $\mathbf{v}_i$).
This is because we expect the basis user embeddings to have different spaces with the user embeddings for ranking to better adapt to the recall task.
The basis user embeddings are further synthesized into a unified user embedding $\mathbf{u}_{re}$ for recall by $\mathbf{u}_{re}=\sum_{i=1}^M \alpha_i \mathbf{v}_i$.
We use a news encoder that is shared with the ranking part to obtain the embedding $\mathbf{r}^c$ of each candidate news $D^c$ in the news pool.
The final recall relevance score $\hat{y}_{re}$ between user interest and candidate news is computed by $\hat{y}_{re}=\mathbf{u}_{re} \cdot \mathbf{r}^c$.

\subsection{Model Training}

Then we introduce the model training details of \textit{UniRec}.
We use a two-stage model training strategy to first learn the ranking part and then learn the recall part.
Following prior works~\cite{huang2013learning,wu2019npa,wu2019nrms}, we use negative sampling techniques to construct samples for contrastive  model learning~\cite{oord2018representation}.
For learning the ranking part, we use clicked news in each impression as positive samples, and we randomly sample $K$ non-clicked news that are displayed in the same impression as negative samples.
The loss function is formulated as follows:
\begin{equation}
    \mathcal{L}_{ra}=-\log {\Big[}\frac{\exp(\hat{y}^+_{ra})}{\exp(\hat{y}^+_{ra})+\sum_{i=1}^K\exp(\hat{y}^{i-}_{ra})}{\Big]},
\end{equation}
where $\hat{y}^+_{ra}$ and $\hat{y}^{i-}_{ra}$ denote the predicted click scores of a positive sample and the corresponding $i$-th negative sample, respectively.
By optimizing this loss function, the parameters of news and user encoders can be tuned.
Motivated by~\cite{ying2018graph}, we fix the news encoder after the ranking model converges.
Then, to learn the recall part, we also use clicked news of each user as positive samples, while we randomly select $T$ non-clicked news from the entire news set as negative samples, which aims to simulate the news recall scenario.
The loss function for recall part training is as follows:
\begin{equation}
    \mathcal{L}_{re}=-\log{\Big[}\frac{\exp(\hat{y}^+_{re})}{\exp(\hat{y}^+_{re})+\sum_{i=1}^T\exp(\hat{y}^{i-}_{re})}{\Big[},
\end{equation}
where $\hat{y}^+_{re}$ and $\hat{y}^{i-}_{re}$ represent the predicted recall relevance scores of a positive sample and the corresponding $i$-th negative sample, respectively.

However, not all basis user embeddings are relevant to the interests of a user.
Thus, motivated by Principal Component Analysis (PCA), in the test phase we propose to only use the top $P$ basis user embeddings with the highest attention weights to compose the user embedding for recall.
We denote these basis user embeddings as $[\mathbf{v}_{t_1},\mathbf{v}_{t_2}, ..., \mathbf{v}_{t_P}]$.
We re-normalize their attention weights as follows:
\begin{equation}
    \alpha_{t_i}=\frac{\exp(\alpha_{t_i})}{\sum_{j=1}^P \exp(\alpha_{t_j})}.
\end{equation}
The user embedding $\mathbf{u}_{re}$ for recall  is built by $\mathbf{u}_{re}=\sum_{i=1}^P \alpha_{t_i}\mathbf{v}_{t_i}$, which can attend more to the major interests of a user and filter noisy basis user embeddings for better news recall.

\subsection{Complexity Analysis}

We provide some discussions on the computational complexity.
In existing news recommendation methods that conduct recall and ranking with separate models, the computational complexity of learning user embeddings for recall and ranking are both $O(N)$ at least, because they need to encode the entire user behavior sequence.
\textit{UniRec} has the same complexity in learning the user embedding for ranking, but the complexity of deriving the user embedding for recall is reduced to $O(M)$, where $M$ is usually much smaller than $N$.
In addition, the attention network used for synthesizing the user embedding for recall may also be lighter-weight than the user encoder.
Thus, the total computational complexity  can be effectively reduced.

\section{Experiments}\label{sec:Experiments}

\subsection{Dataset and Experimental Settings}

We conduct experiments on a large-scale public dataset named MIND~\cite{wu2020mind} for news recommendation.
It contains news impression logs of 1 million users on Microsoft News in 6 weeks.
The logs in the first five weeks are for training and validation, and the rest logs are for test.
The detailed statistics of MIND are shown in Table~\ref{dataset}.

\begin{table}[h]
\centering

\resizebox{0.48\textwidth}{!}{
\begin{tabular}{lrlr}
\Xhline{1.5pt}
\# Users                & 1,000,000  & \# News                      & 161,013    \\
\# Impressions               & 15,777,377 & \# Click behaviors      & 24,155,470  \\ 
Avg. news title len.& 11.52  & \# Categories &  20  \\ \hline \Xhline{1.5pt}
\end{tabular}
}
\caption{Statistics of the \textit{MIND} dataset.}\label{dataset}
\end{table}

In our experiments, following~\cite{wu2020mind} we use news titles to learn news embeddings.
The number of basis user embeddings is 20, and they are randomly initialized.
The hyperparameter $P$ that controls the number of basis user embeddings for composing the user embedding for recall in the test phase is 5.
The number of negative samples associated with each positive one is 4 and 200 for the ranking and recall tasks, respectively.
Adam~\cite{kingma2014adam} is used as the optimizer.
The batch size is 32.
These hyperparamters are selected on the validation set.
Following~\cite{wu2020mind}, we use AUC, MRR, nDCG@5 and nDCG@10 to evaluate news ranking performance.
In addition, we use recall rate of the top 100, 200, 500 and 1000 ranked news to evaluate news recall performance.
We repeat every experiment 5 times.

\begin{table*}[h]
\centering

\begin{tabular}{l|cccc}
\Xhline{1.5pt}
 \textbf{Methods} & \textbf{AUC}           &  \textbf{MRR}            & \textbf{nDCG@5}          &\textbf{nDCG@10}  \\ \hline 
EBNR                                 & 66.22$\pm$0.17  & 31.97$\pm$0.14  & 34.89$\pm$0.17  & 40.49$\pm$0.19   \\
DKN                                  & 65.61$\pm$0.20  & 31.58$\pm$0.17  & 34.32$\pm$0.19  & 40.04$\pm$0.22   \\
NPA                                  & 67.62$\pm$0.14  & 32.69$\pm$0.13  & 35.52$\pm$0.15 & 41.33$\pm$0.17   \\
NAML                                & 67.45$\pm$0.12  & 32.48$\pm$0.09  & 35.39$\pm$0.10  & 41.19$\pm$0.14   \\
NRMS                                & 68.24$\pm$0.09  & 33.38$\pm$0.10  & 36.34$\pm$0.10  & 42.12$\pm$0.13   \\ \hline
UniRec      & \textbf{68.41}$\pm$0.11  & \textbf{33.50}$\pm$0.10  & \textbf{36.47}$\pm$0.12  & \textbf{42.26}$\pm$0.14  \\
\Xhline{1.5pt}
\end{tabular}
\caption{Ranking performance of different methods.}

\label{table.result}
\end{table*}

\begin{table*}[h]
\centering

\begin{tabular}{l|cccc}
\Xhline{1.5pt}
 \textbf{Methods} & \textbf{R@100}           & \textbf{R@200}            & \textbf{R@500}         & \textbf{R@1000} \\ \hline 
YoutubeNet                                     & 1.395$\pm$0.034  & 2.284$\pm$0.039  & 4.171$\pm$0.042  & 6.867$\pm$0.037   \\
Pinnersage                                  & 1.431$\pm$0.020  & 2.340$\pm$0.018  & 4.252$\pm$0.017  & 6.927$\pm$0.019   \\
Octopus                                    & 1.426$\pm$0.026  & 2.392$\pm$0.029  & 4.344$\pm$0.031  & 7.188$\pm$0.029   \\ \hline
UniRec(all)                                 & 1.443$\pm$0.023  & 2.402$\pm$0.027 & 5.022$\pm$0.025  & 8.294$\pm$0.026   \\ 
UniRec(top)      & \textbf{1.516}$\pm$0.026  & \textbf{2.531}$\pm$0.024  & \textbf{5.142}$\pm$0.027  & \textbf{8.485}$\pm$0.026  \\
\Xhline{1.5pt}
\end{tabular}
\caption{Recall performance of different methods.}

\label{table.result2}
\end{table*}

\subsection{Performance Evaluation}

We first compare the ranking performance of \textit{UniRec} with several baseline methods, including:
(1) EBNR~\cite{okura2017embedding}, GRU~\cite{cho2014learning} network for user interest modeling in news recommendation;
(2) DKN~\cite{wang2018dkn}, deep knowledge network for news recommendation;
(3) NPA~\cite{wu2019npa}, news recommendation with personalized attention;
(4) NAML~\cite{wu2019}, news recommendation with attentive multi-view learning;
(5) NRMS~\cite{wu2019nrms}, news recommendation with multi-head self-attention.
The ranking performance of different methods is shown in Table~\ref{table.result}.
We find that \textit{UniRec} outperforms several compared baseline methods like NAML and NPA.
This may be because self-attention has stronger ability in modeling news and user interests.
In addition, \textit{UniRec} also slightly outperforms its basic model NRMS.
This is because \textit{UniRec} can capture the orders of words and behaviors via position embedding.

In the news recall task, we compare \textit{UniRec} with top basis user embeddings (denoted as \textit{UniRec}(top)) with the following baseline methods:
(1) YoutubeNet~\cite{covington2016deep}, using the average of clicked news embeddings for recall;
(2) Pinnersage~\cite{pal2020pinnersage}, an item recall method based on hierarchical clustering;
(3) Octopus~\cite{liu2020octopus}, learning elastic number of user embeddings for item recall;
(4) UniRec(all), a variant of \textit{UniRec} that uses all basis user embeddings to compose the user embedding for recall.
We show the recall performance of different methods in Table~\ref{table.result2}.
We find YoutubeNet is less performant than other recall methods.
This may be because different user behaviors may have different importance in user interest modeling and simply average their embeddings may be suboptimal.
In addition, both \textit{UniRec(top)} and \textit{UniRec(all)} outperform other baseline methods.
This is because our approach can exploit the user interest information inferred from the  ranking module to enhance news recall.
In addition, our approach is a unified model for both recall and ranking, which has better efficiency in online systems than other methods.
Besides, \textit{UniRec(top)} outperforms its variant \textit{UniRec(all)}.
It may be because selecting the basis user embeddings with top attention weights can learn accurate user interest embeddings by attending to major user interests and filtering noisy ones.
The above results validate the effectiveness of our method in both news ranking and recall.

\begin{figure*}[!t]
  \centering
    \includegraphics[width=0.9\linewidth]{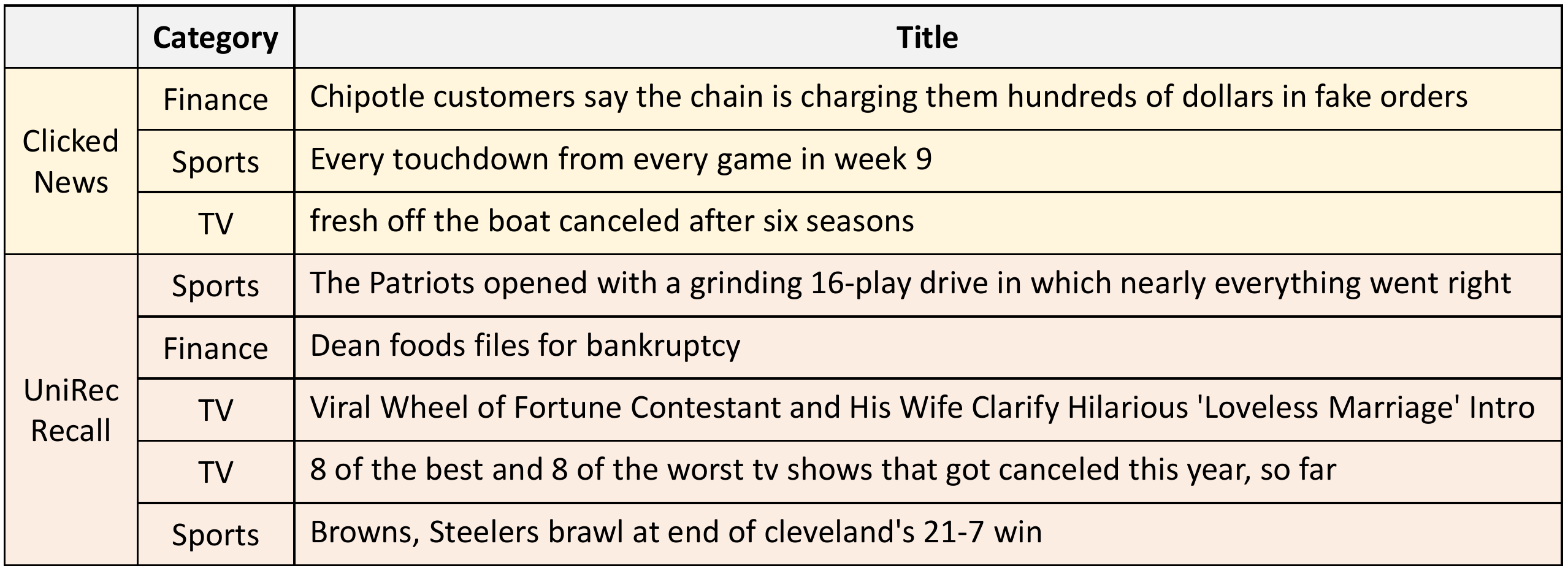}
  \caption{The news clicked by a randomly sampled user and the top news recalled by \textit{UniRec}.}
  \label{fig.case}
\end{figure*}

\subsection{Case Study}

We verify the effectiveness of \textit{UniRec} in news recall via several case studies. 
Fig.~\ref{fig.case} shows the clicked news of a random user and several top news recalled by \textit{UniRec}.
From the user's clicked news, we can infer that this user may be interested in finance, sports and TV shows.
We find the recall result of \textit{UniRec} covers user interest categories of clicked news, but also keeps some diversity with them.
It shows that \textit{UniRec} can generate accurate and diverse personalized news recall results.

\begin{figure}[t]
  \centering
    \includegraphics[width=0.9\linewidth]{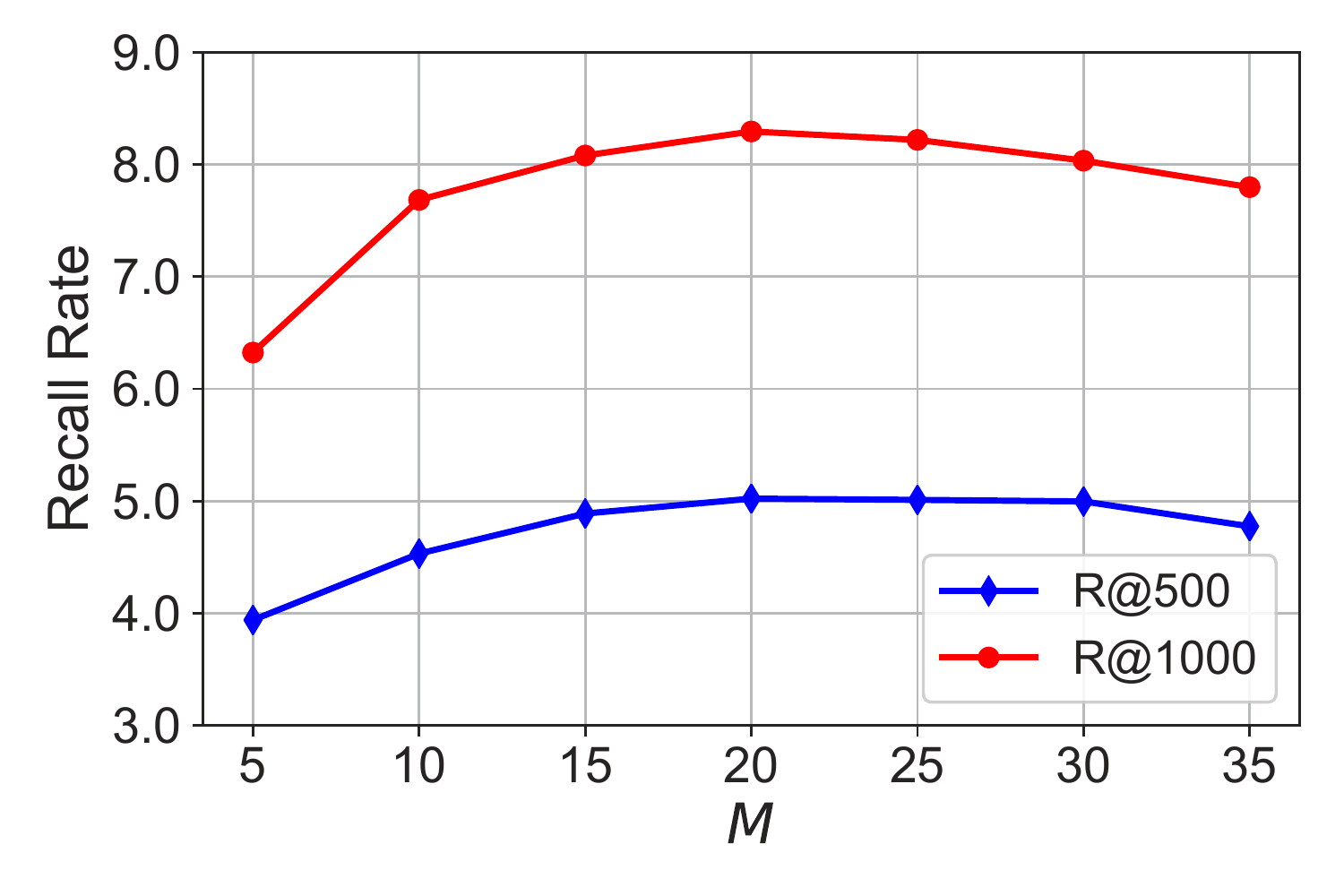}
  \caption{Influence of the basis user embedding number.}
  \label{fig.m}
\end{figure}
\begin{figure}[t]
  \centering
    \includegraphics[width=0.9\linewidth]{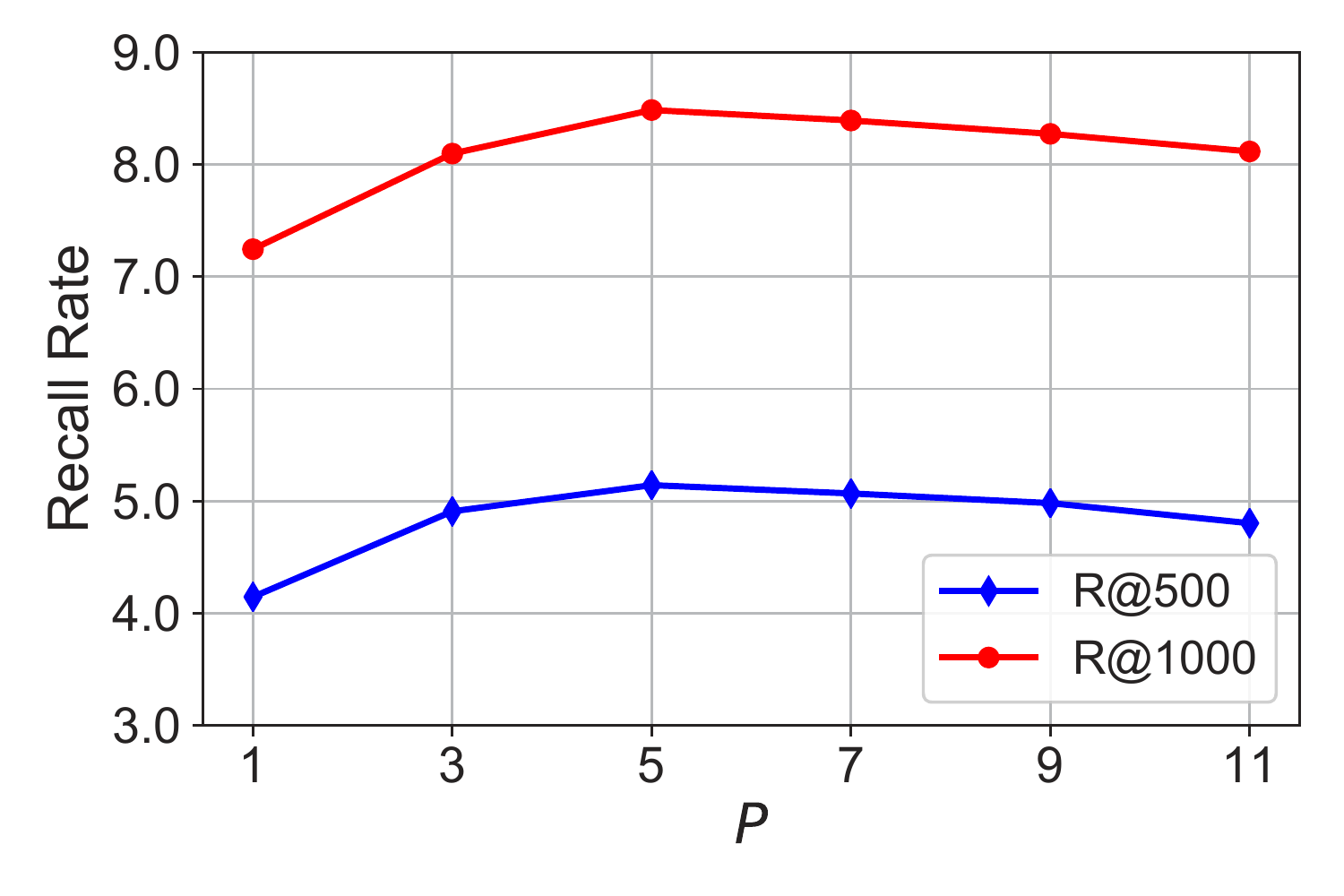}
  \caption{Influence of the hyperparameter $P$.}
  \label{fig.p}
\end{figure}

\subsection{Hyperparameter Analysis}

Finally, we study the influence of two important hyperparameters in our \textit{UniRec} method, including the total number $M$ of basis user embeddings and the number $P$ of basis user embeddings for composing the user embeddings for recall.
We first set $P=M$ and tune the value of $M$.
The recall performance is shown in Fig.~\ref{fig.m}.
We find the performance is suboptimal when $M$ is too small, which may be due to the diverse user interests cannot be covered by a few basis user embeddings.
However, the performance also descends when $M$ is large.
This may be because it is difficult to accurately select informative basis user embeddings for user interest modeling.
In addition, the computation and memory costs also increase.
Thus, we set $M$ to a medium value (i.e., 20) that yields the best performance.
We then tune the value of $P$ under $M=20$.
The results are shown in Fig.~\ref{fig.p}.
We find the performance is suboptimal when $P$ is very small.
This is intuitive because the user interests cannot be fully covered.
However, the performance also declines when $P$ is relatively large.
This may be because basis user embeddings with relatively low attention weights are redundant or even noisy for user interest modeling.
Thus, we choose to use 5 basis user embeddings to compose the  user embedding for recall.
\section{Conclusion}\label{sec:Conclusion}

In this paper, we present a unified approach for  recall and ranking in news recommendation.
In our method, we first infer a user embedding for ranking from historical news click behaviors via a user encoder model.
Then we derive a user embedding for recall from the obtained user embedding for ranking by regarding it as attention query to select a set of basis user embeddings that encode different general user interests.
Extensive experiments on a benchmark dataset validate the effectiveness of our approach in both news ranking and recall.

\section*{Acknowledgments}
This work was supported by the National Natural Science Foundation of China under Grant numbers U1936208, U1936216, and 61862002, and the research initiation project of Zhejiang Lab (No. 2020LC0PI01).

\bibliography{main}
\bibliographystyle{acl_natbib}

\end{document}